\documentclass[preprint,showpacs,aps,pre]{revtex4-1}
\usepackage{epsfig}
\usepackage{psfrag}
\usepackage{graphicx}
\usepackage{subfig}
\usepackage{epstopdf}
\usepackage{graphics}
\usepackage{amsmath}
\usepackage[utf8]{inputenc}
\usepackage{color,graphics}

\usepackage{wrapfig} 
\usepackage{float} 

\begin{document}


\title{Modeling of relativistic ion-acoustic waves in ultra-degenerate plasmas} 

\author{Fernando Haas}
\affiliation{Instituto de F\'{\i}sica, Universidade Federal do Rio Grande do Sul, Av. Bento Gon\c{c}alves 9500, 91501-970 Porto Alegre, RS, Brasil}

\begin{abstract}
We consider the relativistic ion-acoustic mode in a plasma composed by cold ions and an ultra-degenerate electron gas, described the relativistic Vlasov-Poisson system. A critical examination of popular fluid models for relativistic ion-acoustic waves is provided, comparing kinetic and hydrodynamic results. The kinetic linear dispersion relation is shown to be reproduced by the rigorous relativistic hydrodynamic equations with Chandrasekhar's equation of state.
\end{abstract}

\maketitle

\section{Introduction}
Recently, there has been a lot of interest on relativistic  waves in degenerate plasmas, described by hydrodynamic equations \cite{Ali, Haas1, Hussain, Irfan, Kalejahi, Masood, McKerr1, McKerr2, Rahman}. The examination of the literature shows sometimes the use of questionable options, namely: choosing to work with the proper number density (which is the number density in a local reference frame where the fluid is at rest) in the equation of state, while using the laboratory number density in the continuity equation, with the same symbol for both objects; the use of relativistic equations of state inside otherwise non-relativistic fluid equations; the use of non-relativistic equations of state inside otherwise relativistic fluid equations; covariant or non-covariant form of the pressure term, including or not the respective time-derivative; taking into account or not, the relativistic mass increase due to thermal effects. Fortunately, similar criticisms have been already made \cite{Berezhiani, Lee, Lontano, Saberian}. However, it is the time to stress once again the need of a more firm ground on the choice of relativistic fluid equations. A good way to proceed, is by comparison between fluid models and relativistic kinetic theory, which is by definition more general than the macroscopic approaches. The right choice of hydrodynamic equations is important, for instance, for the development of exact nonlinear waves and solitons \cite{Infeld}, hardly accessible by means of kinetic theory. 

In this context, here we will derive the linear dispersion relation for ion-acoustic waves in a deep degenerate plasma composed by electrons and cold, massive ions, using the relativistic 
Vlasov-Poisson system. In a first approximation, quantum recoil will be neglected, not only because of simplicity, but also because quantum diffraction effects contribute with a $\sim k^4$ term in the wave dispersion \cite{Haas}, where $k$ is the wave-number. As far as the ion-acoustic velocity $c_s$ is concerned, we can keep ourselves with the lowest-order term, since $\omega \approx c_s k$, where $\omega$ is the wave frequency. 

The calculation of the plasma (longitudinal and transverse) response in fully degenerate relativistic plasma has been made by Jancovici \cite{Jancovici}, using the dielectric formalism of a quasi-boson Hamiltonian approximation. Jancovici's result generalizes the non-relativistic expression found by Lindhard \cite{Lindhard}. In our case, for the ion-acoustic wave, the static ($\omega \approx 0$) limit of the electronic response will be sufficient. The static response is also relevant for Kohn's anomaly and Friedel oscillations \cite{Kohn} and, in the case of the transverse response, for the magnetic susceptibility. Moreover, it has a deep influence on a possible attractive quantum force between ions in plasmas \cite{Shukla1}, not yet empirically tested, and subjected to some theoretical controversy \cite{Shukla2, Tyshetskiy}. An excellent historical overview of the literature on dispersion relations for relativistic degenerate plasma is given in \cite{Melrose}, chapter IX; see also \cite{Kowalenko}. Our aim is to verify the agreement between relativistic kinetic and fluid theories for ion-acoustic waves in degenerate plasmas, as put forward in recent works \cite{Haas1, McKerr1, McKerr2}.

This work is organized as follows. In Section II, we review the derivation of ion-acoustic waves in terms of the relativistic Vlasov-Poisson system. Section III makes the comparison with the results from the (rigorous) hydrodynamic model. Section IV shows our conclusions. 
 
\section{Ion-acoustic waves in a relativistic degenerate plasma: kinetic theory}
The relativistic Vlasov-Poisson system for an electron-ion plasma is given by 
\begin{eqnarray}
\frac{\partial f_e}{\partial t} + \frac{\bf p}{\gamma_e m_e}\cdot\nabla f_e - e {\bf E}\cdot\frac{\partial f_e}{\partial{\bf p}} &=& 0 \,, \\ 
\label{fi}
\frac{\partial f_i}{\partial t} + \frac{\bf p}{\gamma_i m_i}\cdot\nabla f_i + e {\bf E}\cdot\frac{\partial f_i}{\partial{\bf p}} &=& 0 \,, \\
\nabla\cdot{\bf E} = \frac{e}{\varepsilon_0}\,\int d^3 p\, (f_i - f_e)  \,,
\end{eqnarray}
where $f_{e,i} = f_{e,i}({\bf r},{\bf p},t)$ denote the electron (e) and ion (i) phase space probability distribution functions, 
$m_{e,i}$ are the electron and ion masses, $\gamma_{e,i} = \sqrt{1 + p^2/(m_{e,i}^2 c^2)}$ are the relativistic gamma factors, ${\bf E}$ is the electric field, $e$ is the elementary charge (for simplicity, ions are supposed to be singly ionized) and $\varepsilon_0$ is the vacuum permittivity. The integrals are performed in the whole momentum space. By assumption, we also have $m_i \gg m_e$. 

The equilibrium state is ${\bf E} = 0$, with a cold ionic distribution $f_{i}^0 = n_0 \delta({\bf p})$ and an electronic distribution $f_e = f_{e}^{0}({\bf p})$, normalized so that $\int d^3 p f_{e}^{0}({\bf p}) = n_0$.  Assuming, as usual, plane wave perturbations $\sim \exp[i({\bf k}\cdot{\bf r} - \omega t)]$, we get the linear dispersion relation 
\begin{equation}
\label{rdr}
\epsilon({\bf k},\omega) = 1 - \frac{\omega_{pi}^2}{\omega^2} - \frac{\omega_{pe}^2}{n_0}\,\int \frac{d^3 p}{\gamma_{e}^3} \frac{f_{e}^{0}({\bf p})}{[\omega - {\bf k}\cdot{\bf p}/(\gamma_e m_e)]^2}\,\left(1 + \frac{p^2}{m_{e}^2 c^2}\,\sin^{2}\theta\right) = 0 \,, 
\end{equation}
where ${\bf k} = k \hat{z}, {\bf p} = p\,(\cos\phi\sin\theta,\sin\phi\sin\theta,\cos\theta)$ and $\omega_{p\,e,i} = [n_0 e^2/(m_{e,i}\varepsilon_0)]^{1/2}$. Naturally, for a cold ion distribution the non-relativistic approximations for ions would have been sufficient. Nevertheless, the ionic force equation (\ref{fi}) was keep relativistic with an eye on alternative applications. 

Assume a deep degenerate electronic distribution function given by 
\begin{equation}
f_{e}^0 = 
     \begin{cases}
       3 n_0/(4 \pi p_F^3) \,, &\quad p < p_F \,, \\
       0 \,, &\quad p > p_F  \,, 
     \end{cases}
		\label{fd}
\end{equation}
where $p_F = \hbar\,(3 \pi^2 n_0)^{1/3}$ is the Fermi momentum and $\hbar = h/(2\pi)$ is the reduced Planck constant.

Evaluating the electrons integral, we find the real part of the longitudinal dielectric function given by 
\begin{equation}
\epsilon({\bf k}, \omega) = 1 - \frac{\omega_{pi}^2}{\omega^2} + \frac{3\,\omega_{pe}^2 \sqrt{1 + \xi_0^2}}{c^2 k^2 \xi_0^2}\left(1 - 
\frac{\omega}{2\,k v_F}\,\ln\left|\frac{\omega + k v_F}{\omega - k v_F}\right|\right) = 0 \,,
\label{jan}
\end{equation}
where $v_F = p_F/(\gamma_F m_e)$ is the electrons Fermi velocity in terms of $\gamma_F = \sqrt{1 + \xi_0^2}$ and  $\xi_0 = p_F/(m_e c)$ is the relativistic parameter. We are ignoring the imaginary contribution (collisionless damping) for the moment. The dispersion relation (\ref{jan}) was first obtained by Jancovici \cite{Jancovici}, using a quasi-boson Hamiltonian approximation, and without accounting for the ion dynamics or, equivalently,  taking $m_i/m_e \rightarrow \infty, \omega_{pi} \rightarrow 0$. The longitudinal plasma response including the ions motion was later reported in \cite{Delsante}.  

For ion-acoustic waves in degenerate plasma by definition we have 
\begin{equation}
\label{ineq}
v_{Fi} \ll \omega/k \ll v_F \,, 
\end{equation}
where $v_{Fi}$ is the ions Fermi velocity. For simplicity, the small correction arising from the ions Fermi temperature (or ions thermodynamic temperature in the case of non-degenerate ions) will be essentially ignored in what follows. Notice the similarity with ion-acoustic waves in classical plasma, with the thermal electron and ion velocities replacing respectively the electron and ion Fermi velocities. 

\begin{figure}
   \centerline{\includegraphics[height=7cm,width=13cm]{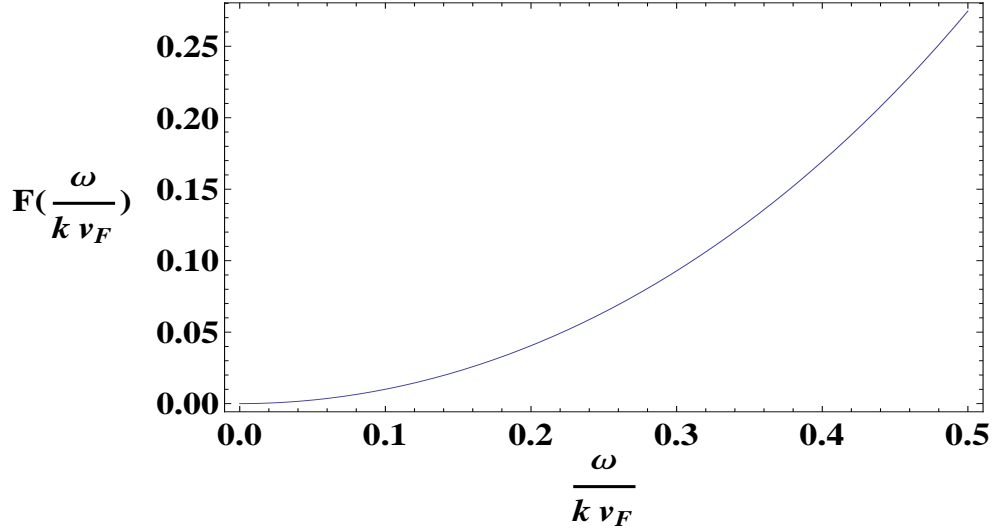}}
  \caption{Function $F[\omega/(k\,v_F)]$ defined in Eq. (\ref{ff}), responsible for the dynamical part of the electron response in the longitudinal dielectric function (\ref{jan}).}
\label{fig1}
\end{figure}

In view of Eq. (\ref{ineq}), the static electronic response ($\omega \approx 0$) is sufficient for ion-acoustic waves, as can be seen from the behavior of the function $F[\omega/(k v_F)]$ present in the last term of the dielectric constant (\ref{jan}), 
\begin{equation}
\label{ff}
F\left(\frac{\omega}{k\,v_F}\right) = \frac{\omega}{2\,k\,v_F}\,\ln\left|\frac{1 + \frac{\omega}{k\,v_F}}{1 - \frac{\omega}{k\,v_F}}\right| \,,
\end{equation}
shown in Fig. \ref{fig1}, ignoring the small imaginary part of the frequency (as later discussed below). As apparent, the smallness of the function $F[\omega/(k v_F)]$ for slow modes so that $\omega/k \ll v_F$ shows that the dynamic electronic response is not necessary. 

Therefore, we get 
\begin{equation}
\label{kin}
\epsilon({\bf k}, \omega) \approx 1 - \frac{\omega_{pi}^2}{\omega^2} + \frac{3\,\omega_{pe}^2 \sqrt{1 + \xi_0^2}}{c^2 k^2 \xi_0^2} = 0 \,,
\end{equation}
so that 
\begin{equation}
\label{iaw}
\omega^2 = \frac{c_s^2 k^2}{1 + c_s^2 k^2/\omega_{pi}^2} \,,
\end{equation}
where
\begin{equation}
\label{cs}
c_s^2 = \frac{p_F^2}{3\,m_e m_i \sqrt{1 + \xi_0^2}}
\end{equation}
provides the definition of the ion-acoustic velocity $c_s$. The dispersion relation (\ref{iaw}) encompasses both non-relativistic and ultra-relativistic regimes. As can be checked, one always has $(c_s/v_F)^2 = m_e \gamma_F/(3 m_i) \ll 1$ as long as $n_0 \ll 10^{47}\, {\rm m}^{-3}$, in accordance with Eq. (\ref{ineq}). 

\begin{figure}
   \centerline{\includegraphics[height=7cm,width=13cm]{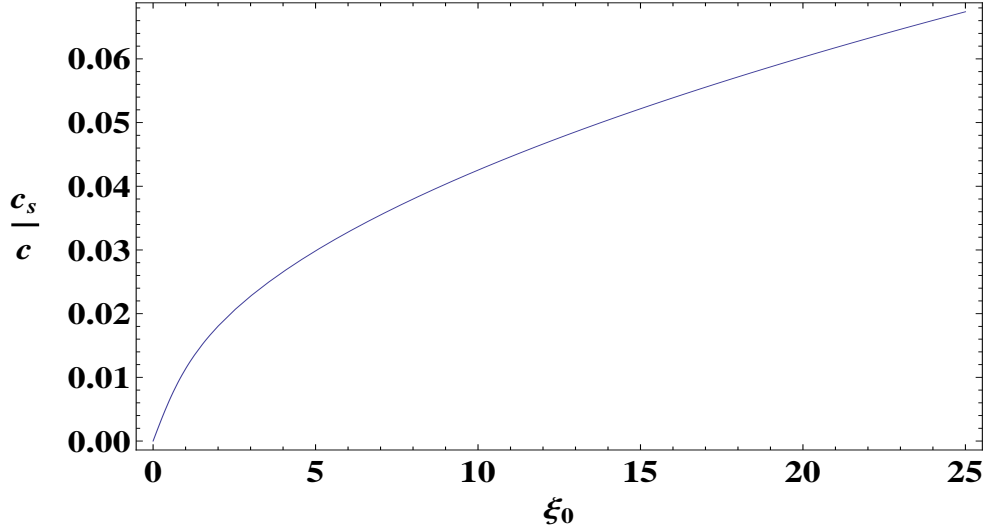}}
  \caption{Normalized relativistic-degenerate ion-acoustic velocity from Eq. (\ref{css}), in terms of the relativistic parameter $\xi_0 = p_F/(m_e\,c)$, for hydrogen plasma parameters.}
\label{fig2}
\end{figure}

Alternatively, one might re-express Eq. (\ref{cs}) as 
\begin{equation}
\label{css}
\frac{c_s}{c} = \left(\frac{m_e\,\xi_0^2}{3\,m_i\,\sqrt{1 + \xi_0^2}}\right)^{1/2} \,,
\end{equation}
where the right-hand side is now a function of the density $n_0$ contained in the relativistic parameter $\xi_0$. The result is shown in Fig. \ref{fig2}, for hydrogen plasma. For a very dense white dwarf with $n_0 = 10^{40}\,{\rm m}^{-3}$, one will have $\xi_0 = 25.73$. Even in this extreme case, the relativistic-degenerate ion-acoustic velocity remains much smaller than the light velocity.

We could examine the assumption of deep degeneracy, which requires $T_F \ll T$, where $T_F = (\gamma_F - 1)\,m_e\,c^2/\kappa_B$ is the Fermi temperature ($\kappa_B$ is the Boltzmann constant) and $T$ is the electrons temperature. For a white dwarf with central temperature $T = 10^7 \,{\rm K}$, one has $T_F > T$ provided $n_0 > 1.15 \times 10^{32}\,{\rm m}^{-3}$, or equivalently 
$\xi_0 > 0.06$. On the other hand, replacing the electron mass by the proton mass whenever necessary and supposing the same ionic temperature $T = 10^7 \,{\rm K}$, one would have degenerate ions for much larger densities, or $n_0 > 9.00 \times 10^{36} \,{\rm m}^{-3}$. 

The imaginary part of the longitudinal dielectric function was analyzed in \cite{Delsante}. Essentially, the result for ultra-degenerate electrons is that collisionless damping is very small, provided $p_F/(m_i c) \ll 1$, a condition fairly well satisfied except for huge densities of the order $n_0 > 10^{45} \,{\rm m}^{-3}$, or $\xi_0 > 10^3$. A general treatment about the imaginary part of the longitudinal dielectric response in non-degenerate relativistic plasma can be found in \cite{Eliasson}. 

\begin{figure}
   \centerline{\includegraphics[height=7cm,width=13cm]{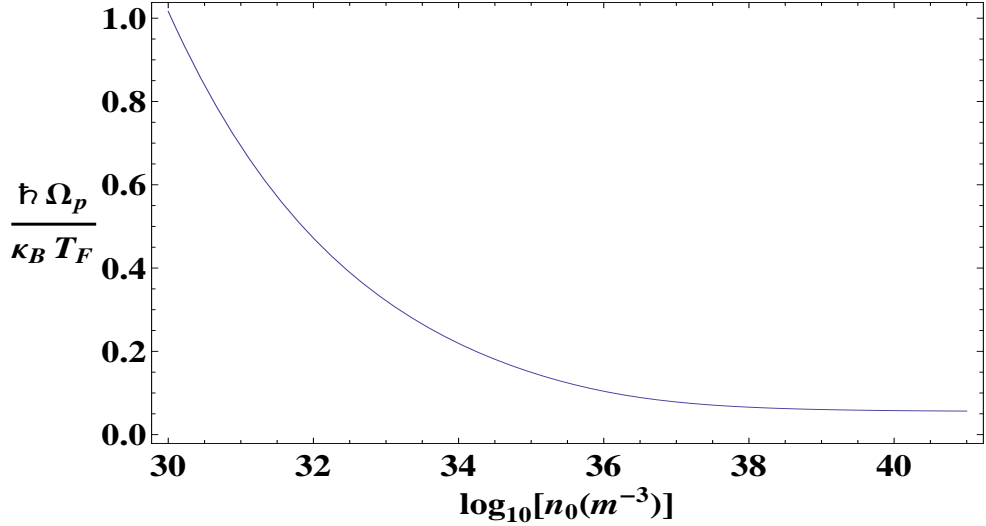}}
  \caption{Ratio between the plasmon energy $\hbar\Omega_p$ and the Fermi energy $\kappa_B T_F$ as a function of density using S.I. units, for hydrogen plasma parameters.}
\label{fig3}
\end{figure}

When reaching enormous densities, one would be faced with the issue of collisionless pair creation, whenever $\hbar\Omega_p > 2 \,m_e \,c^2$. Here $\Omega_p = [n_0 e^2/(\gamma_F m_e \varepsilon_0)]^{1/2}$ is the relativistic plasmon frequency taking into account the relativistic mass increase due to the Fermi momentum. One find that pair creation is avoided provided $n_0 < 2.7 \times 10^{40}\,{\rm m}^{-3}$, or $\xi_0 < 35.83$. In the same trend, positrons can not be excited from the Fermi sea since the plasmon energy $\hbar\Omega_p$ is far less than the Fermi energy, for dense plasmas. This can be seen in Fig. \ref{fig3}, also showing that the plasma becomes more ideal (colisionless) as the density increases. In the ultra-relativistic limit of ultra-high densities, one approaches the asymptotic value $\hbar\Omega_p/(\kappa_B T_F) \rightarrow 2 \sqrt{\alpha/(3\pi)} = 0.06$, where $\alpha = e^2/(4\pi\varepsilon_0\hbar c)$ is the fine-structure constant. 

\section{Ion-acoustic waves in a relativistic degenerate plasma: fluid  theory}

Consider the fluid model for relativistic degenerate electrostatic plasma put forward in \cite{Haas1, McKerr1, McKerr2}, 
\begin{eqnarray}
\frac{\partial}{\partial t}(\bar{\gamma}_\alpha n_\alpha) + \nabla\cdot(\bar{\gamma}_\alpha n_\alpha {\bf u}_\alpha) &=& 0 \,, \quad \alpha = e,i \, \\
\label{ei}
H m_e \Bigl(\frac{\partial}{\partial t} + {\bf u}_e\cdot\nabla\Bigr)(\bar{\gamma}_e {\bf u}_e) &=& - \frac{\bar{\gamma}_e}{n_e} \Bigl(\nabla + \frac{{\bf u}_e}{c^2}\,\frac{\partial}{\partial t}\Bigr)P - e{\bf E} \,,\\
m_i \Bigl(\frac{\partial}{\partial t} + {\bf u}_i\cdot\nabla\Bigr)(\bar{\gamma}_i {\bf u}_i) &=& e{\bf E} \,,\\
\nabla\cdot{\bf E} = \frac{e}{\varepsilon_0} (\bar{\gamma}_i n_i \! &-& \! \bar{\gamma}_e n_e) \,,
\end{eqnarray}
where $n_{e,i}$ and ${\bf u}_{e,i}$ are the electron (ion) proper number densities and velocity fields, $\bar{\gamma}_{e,i} = 1/\sqrt{1 - u_{\alpha}^2/c^2}$ the corresponding relativistic factors, $H$ an enthalpy-like quantity related to relativistic mass increase due to the incoherent bulk motion of the electrons, and $P$ the electrons pressure, to be specified by the appropriate equation of state.  The remaining symbols have the same meaning as before. In the same way as in the kinetic treatment, ions are assumed to be cold. Actually in the present case the ion equations could have been completely classical, but have been written in relativistic form for the sake of generality. 

Notice that the fluid equations involve the proper and not the laboratory densities $N_{e,i} = \gamma_{e,i} n_{e,i}$. Obviously, one could formulate the basic equations in terms of $N_{e,i}$, but in this case for the sake of coherence one would be obliged to insert gamma factors inside the equations of state - a cumbersome circumstance in our view. In addition, we note the covariant form of the pressure term in the electron force equation (\ref{ei}), containing a time-derivative. For linear waves with zero equilibrium velocity, such a term has no role, but the same can not be assured for relativistic speeds and/or fast temporal variations of the fluid pressure. 

In the present ultra-degenerate case, it is indicated to consider the Chandrasekhar equation of state \cite{Chandrasekhar, Oppenheimer}. Hence we set 
\begin{equation}
\frac{P}{n_0 m_e c^2} = \frac{1}{8\,\xi_0^3}\,\left[\xi (2 \xi^2 -3) \sqrt{1 + \xi^2} + 3\, {\rm senh}^{-1}\xi\right] \,,
\end{equation}
where $\xi_0 = p_F/(m_e c)$ is the relativistic parameter as before and $\xi = \xi_0 (n_e/n_0)^{1/3}$. Moreover, one have the enthalpy-like quantity $H = \sqrt{1 + \xi^2}$, which can be related to the mass-energy density. For a recent detailed derivation, see \cite{Haas1}. The equation of state is fully consistent with the isotropic three-dimensional equilibrium (\ref{fd}). It can also be understood as the isothermal equation of state of the Fermi gas, in the limit of zero temperature. Notice that it is not an adiabatic equation of state, which would be inappropriate for ion-acoustic waves. The same happens for classical, non-degenerate, non-relativistic plasmas: isothermal equation of state is the right choice for ion-acoustic waves; adiabatic equation of state holds for Langmuir waves \cite{Krall}. The choice of equation of state for relativistic plasma can be a subtle problem, see \cite{Pegoraro} for the analysis of Langmuir and ion-acoustic waves in the non-degenerate case. 

For the propagation of ion-acoustic waves, the electrons inertia can be entirely neglected, since in the left-hand side of Eq. (\ref{ei}) one has $H m_e << m_i$ as far as $n_e \ll 3.62 \times 10^{45}\,{\rm m}^{-3}$. Therefore, for inertialess electrons, linearizing around $n_{e,i} = n_0, {\bf u}_{e,i} = 0, {\bf E} = 0$ and proceeding as before, the result is 
\begin{equation}
\label{drf}
\epsilon({\bf k},\omega) = 1 - \frac{\omega_{pi}^2}{\omega^2} + \frac{m_e \omega_{pe}^2}{k^2 (dP/dn_e)_0} = 0 \,.
\end{equation}
Moreover, 
\begin{equation}
\frac{dP}{dn_e} = \frac{m_e c^2 \xi^2}{3\,\sqrt{1 + \xi^2}} \quad \Rightarrow \quad \left(\frac{dP}{dn_e}\right)_0 = \frac{p_F^2}{3\,m_e\sqrt{1 + \xi_0^2}}   \,,
\end{equation}
which can be used to show the complete equivalence between (\ref{drf}) and the kinetic longitudinal dielectric function (\ref{kin}). Therefore, the proposed fluid model satisfies the necessary condition, of agreement with the microscopic (kinetic) theory. 

\section{Conclusion}
The comparison between the kinetic and rigorous fluid models of ion-acoustic waves in a relativistic degenerate plasma has been made. As far as the real part of the longitudinal dielectric function is concerned, the agreement has been found to be exact. The spirit of this approach should be promoted, so as to provide a more clear justification of relativistic plasma hydrodynamics models. Applications can be easily pursued e.g. in the case of laser-plasma interactions in the high-energy density regimes. Extension to problems involving significant quantum diffraction is ongoing and will be reported elsewhere.

{\bf Acknowledgments}: the author acknowledges  CNPq  (Conselho  Nacional  de  Desenvolvimento Cient\'{\i}fico e Tecnol\'ogico) for  financial support.

\end{document}